\documentclass[twocolumn,aps,prd]{revtex4}
\usepackage{amsfonts}
\usepackage{amsmath}
\usepackage{amssymb}
\usepackage{graphicx}

\newcommand{\be}{\begin{equation}}
\newcommand{\ee}{\end{equation}}
\newcommand{\ben}{\begin{eqnarray}}
\newcommand{\een}{\end{eqnarray}}

\begin{document}

\title{Lorentz violating effects on a quantized two-level system}
\author{M. M. Ferreira Jr.}
\affiliation{Universidade Federal do Maranh\~ao (UFMA), Departamento de F\'\i sica \\
Campus Universit\'ario do Bacanga, 65085-580, S\~ao Lu\'\i s, Maranh\~ao,
Brazil, e-mail: manojr@ufma.br}
\author{A.R. Gomes}
\affiliation{Departamento de Ci\^encias Exatas, Centro Federal de Educa\c c\~ao
Tecnol\'ogica do Maranh\~ao, 65025-001 S\~ao Lu\'\i s, Maranh\~ao, Brazil,
e-mail: {argomes@pq.cnpq.br}}

\begin{abstract}
In this work, we consider the effects of the Lorentz-violating (LV)\ term $%
v_{\mu }\overline{\psi }\gamma ^{\mu }\psi $ belonging to the fermion sector
of the extended standard model on the dynamics of a quantum two-level
system. We examine how its non-relativistic counterpart, ${(\mathbf{p}-e%
\mathbf{A})\cdot }\mathbf{v}/{m_{e},}$ affects the Rabi oscillations of a
two-level atom coupled with a quantum cavity electromagnetic field. Taking
an initial coherent field state in a resonant cavity, it was found that the
LV\ background increases the Rabi frequency and the time interval between
collapses and revivals of the population inversion function. It was found
that initial field states with low mean number of photons are better probes
in order to establish more stringent upper bounds on the background
magnitude. In particular, for an initial vacuum state in the cavity the
upper limit $\text{v}_{x}<10^{-10}eV$ was attained.
\end{abstract}

\maketitle

%\pacs{xxx,yyy,zzz.}

%%%%%%%%%%%%%%%%%%%%%%%%%%%%%%%%%%%

\section{introduction}

In the latest years, Lorentz violation (LV)\ in physical systems has been
investigated in connection with a possible breakdown of this symmetry at the
Planck scale. Since the demonstration that spontaneous breaking of Lorentz
symmetry may occur in the context of string theory \cite{Samuel}, small
violations of Lorentz covariance in low-energy systems have been searched as
a remanent effect of LV at the Planck scale. Naturally, this is a relevant
issue, once the Planck scale physics is entirely unknown yet. Nowadays, most
LV\ investigations have been conducted into the framework of the Standard
Model Extension (SME) \cite{Colladay}, wherein LV is incorporated in all
sectors of interaction and governed by tensor coefficients generated as
vacuum expectation values of tensor quantities of the original symmetric
theory. In such a model, Lorentz breaking takes place only in the particle
frame, where such coefficients behave as a set of numbers. In the observer
frame, the Lorentz covariance remains valid.

In the framework of the SME, LV has been intensively investigated concerning
mainly the photon \cite{Photon} and fermion sectors with many different
purposes, involving radiative corrections \cite{Radiative}, topological
effects \cite{Topological}, CPT probing experiments \cite{CPT}, hydrogen
spectrum \cite{Hydrogen}, and general aspects \cite{general}. Atomic and
optical systems \cite{Optical} (including resonant cavities) have been used
as a laboratory to test the limits of Lorentz covariance, implying stringent
bounds on the LV coefficients.

The fermion Lagrangian of the SME includes two Lorentz and CPT-odd terms, $%
v_{\mu }\overline{\psi }\gamma ^{\mu }\psi ,b_{\mu }\overline{\psi }\gamma
_{5}\gamma ^{\mu }\psi ,$ with $v_{\mu },b_{\mu }$ being the LV coefficients
generated as vacuum expectation values of tensor quantities belonging to the
underlying high energy theory. In a very recent work, it was analyzed the
effect of these two terms on a semi-classical two-level system \cite{Semi},
being observed that they yield alterations on its dynamics, inducing
sensitive modifications on the usual inversion population function and
quantum transitions even in the absence of an external electromagnetic field.

In the present work we consider the effect of the LV term $v_{\mu }\bar{\psi}%
\gamma ^{\mu }\psi $ on a quantum atomic system. The starting point is the
extended Lagrangian 
\begin{equation}
\mathcal{L}%
%TCIMACRO{\U{b4}}%
%BeginExpansion
{\acute{}}%
%EndExpansion
\mathcal{=L}_{Dirac}-v_{\mu }\overline{\psi }\gamma ^{\mu }\psi   \label{L1}
\end{equation}%
where $\mathcal{L}_{Dirac}$ is the usual Dirac Lagrangian ($\mathcal{L}%
_{Dirac}=\frac{1}{2}i\overline{\psi }\gamma ^{\mu }\overleftrightarrow{%
D }_{\mu }\psi -m_{e}\overline{\psi }\psi )$. It yields the following
nonrelativistic Hamiltonian 
\begin{equation}
H=H_{Pauli}+\biggl[-\frac{(\mathbf{p}-e\mathbf{A})\cdot \mathbf{v}}{m_{e}}%
\biggr].
\end{equation}%
The purpose here is to examine the effects implied by the Lorentz-violating
Hamiltonian in the Rabi nutation of a single atom in the vacuum and in a
weak coherent field established in a resonant cavity. Using the amplitude
coefficient method, the Schr\"{o}dinger equation is taken as starting point
to obtain differential equations for the amplitude coefficients. These
equations govern the dynamics of the two-level system and allow to read the
effects induced by the LV background on it. After some approximation, it is
derived a system of two coupled typical harmonic oscillator differential
equations, whose solution leads to a modified expression for the inversion
population function. It then reveals that the Rabi frequency, the collapse $%
\left( t_{c}\right) $ and revival $\left( t_{r}\right) $ times all increase
with the background magnitude. At the same time, the revival packages become
larger and more distant from each other. Considering that quantum
experiments present a sensitivity of 1 part in $10^{10}$, an upper bound $%
\left( \text{v}_{x}\leq 10^{-10}eV\right) $ for the background is
established.

This paper is outlined as follows. In Sec.\textbf{\ }II, it is presented a
brief resume on the interaction of a two-level system with a quantized
monochromatic field. In Sec. III, the Lorentz violating effects stemming
from the coupling $v_{\mu }\overline{\psi }\gamma ^{\mu }\psi $ on the
two-level system are properly examined by means of the amplitude coefficient
method. Special attention is paid to the LV effects on the Rabi frequency,
collapse and revival times. In Sec. IV, we present our Conclusion and final
remarks.

%%%%%%%%%%%%%%%%%%%%%%%%%%%%%%%%%%%%%%%%%%%%%%%%%%%%%%%%%%%%%%%%%%%%%%%%%%%%%%%

\section{Interaction of a two-level atom with a single mode field}

\label{sect2}

First of all, we review a standard result concerning the interaction
between a two-level atom with a quantized field (in the absence of the
background $\mathbf{v)}$. The two levels are identified as $|a\rangle $ and $%
|b\rangle $ with energy $(1/2)\hbar \omega $ and $-(1/2)\hbar \omega ,$
respectively. This system is described by the Hamiltonian in the Schr\"{o}%
dinger representation (See Ref. \cite{scully}, Chap. 6): 
\begin{equation}
\hat{H}=\hat{H}_{atom}+\hat{H}_{field}+\hat{H_{1}},  \label{H1}
\end{equation}%
where $\hat{H}_{atom}=\hbar \omega _{a}|a\rangle +\hbar \omega _{b}|b\rangle 
$ is the atomic Hamiltonian, \ $\hat{H}_{field}=$ $\sum_{k}\hbar \nu _{k}(%
\hat{a}^{\dagger }\hat{a}+1/2)$ is the radiation field Hamiltonian, $\hat{%
H_{1}}=-e\mathbf{r}\cdot \mathbf{E}$ describes the atom-field interaction in
the dipole approximation, and $\hat{a},\hat{a}^{\dagger }$ are the photon
destruction and creation operators. Here, $\mathbf{r}$ is the position
vector of the electron and $\mathbf{E}$ is the radiation field that
interacts with the atom,
\ben
\mathbf{E=}\sum\limits_{k}E_{0k}(\hat{a}e^{-i\nu
_{k}t}+\hat{a}^{\dagger }e^{i\nu _{k}t})\hat{\epsilon}_{k},
\een
 with $\hat{%
\epsilon}_{k}$ being the polarization vector and $E_{0k}$ being the
amplitude of the mode of frequency $\nu _{k}.$ Such amplitude is given by $%
E_{0k}=\sqrt{\hbar \nu _{k}/(2\epsilon _{0}\mathcal{V})},$ where $\mathcal{V}
$ is the cavity effective volume. This normalization factor is obtained by
equating the Fock states energy with the integral over space of the
expectation value of the electromagnetic energy density. In this work, we
will consider the interaction of a single-mode field of frequency $\nu $
with the two-level atom, so from now on we write $\hat{H}_{field}=\hbar \nu (%
\hat{a}^{\dagger }\hat{a}+1/2)$ and $\mathbf{E=}E_{0}(\hat{a}e^{-i\nu t}+%
\hat{a}^{\dagger }e^{i\nu t})\hat{\epsilon}_{k}.$

Now, the Hamiltonian (\ref{H1}) can be read as $\hat{H}=\hat{H}_{0}+\hat{%
H_{1}},$ where $\hat{H}_{0}=\hat{H}_{atom}+\hat{H}_{field}$ \ plays the role
of the unperturbed interaction and $\hat{H_{1}}=-e\mathbf{r}\cdot \mathbf{E}$
can be viewed as a small perturbation. The approach adopted here is
developed in the interaction picture, wherein the state vectors evolve with $%
\hat{H_{1}}$ whereas the operators evolve with $\hat{H_{0}}$ \cite{foot1}.
In such a picture, the interaction operator $H_{1}$ is to be written as $%
\hat{H_{1}}_{I}(t)=e^{iH_{0S}t/\hbar }\hat{H_{1}}_{S}e^{-iH_{0S}t/\hbar },$
where $\hat{H_{1}}_{S}$ stands for the atom-field interaction in the Schr%
\"{o}dinger picture, in which the operators do not present time dependence
(see also Ref. \cite{loudon}, p. 187). For this reason, the time dependence
of $\hat{\text{ }H_{1}}$ will be dropped out from now on.

In order to evaluate $\hat{H_{1}}$ in a more suitable form, we use the atom
transition operators, $\hat{\sigma}_{ij}=|i\rangle \langle j|,$ where $%
|i\rangle $ represents a complete set of energy eigenstates, so that $1=\sum
|i\rangle \langle i|$. In our two-level case, $|i\rangle =|a\rangle $ or $%
|b\rangle ,$ obviously. Considering it, we obtain: 
\begin{equation}
\hat{H_{1}}=\sum_{ij}g^{ij}\hat{\sigma}_{ij}(\hat{a}+\hat{a}^{\dagger }),
\end{equation}%
where the electric field was evaluated at $t=0$ (due to the choice of the
Schr\"{o}dinger representation), $g^{ij}=-e(\mathbf{P}_{ij}\cdot \hat{%
\epsilon}_{k})E_{0k}/\hslash ,$ and $e\mathbf{P}_{ij}=e\langle i|\mathbf{r}%
|j\rangle $ is the transition matrix element of the electric dipole moment.
Supposing that $\mathbf{P}_{ab}=\mathbf{P}_{ba},g^{ab}=g^{ba}=g,$ the
interaction $\hat{H_{1}}$ takes the form $\hat{H_{1}}=g(\hat{\sigma}_{ab}+%
\hat{\sigma}_{ba})(\hat{a}+\hat{a}^{\dagger }).$ The terms $\hat{\sigma}_{ab}%
\hat{a}^{\dagger }$ and $\hat{\sigma}_{ba}\hat{a}\ $should be neglected.
Indeed, the term $\hat{\sigma}_{ab}\hat{a}^{\dagger }$ induces an atomic
transition from the ground state ($|b\rangle $) to the excited state ($%
|a\rangle )$ while a photon of frequency $\nu $ is emitted. The term $\hat{%
\sigma}_{ba}\hat{a}$ \ implies an atomic transition from the excited state ($%
|a\rangle $) to the ground state ($|b\rangle $) while a photon of frequency $%
\nu $ is absorbed. Both processes do not conserve energy. The exclusion of
the non-conserving energy terms is equivalent to the rotating wave
approximation (RWA). In the semiclassical theory it takes place a similar
fact: the non-resonant terms are neglected. We now introduce the notation: $%
\hat{\sigma}_{_{+}}=\hat{\sigma}_{ab}=|a\rangle \langle b|$, $\hat{\sigma}%
_{_{-}}=\hat{\sigma}_{ba}=|b\rangle \langle a|,$ so that the
energy-conserving Hamiltonian takes the form $\hat{H}=\hat{H}_{0}+\hat{H}_{1}
$, with 
\begin{eqnarray}
\hat{H_{0}} &=&\hbar \nu \hat{a}^{\dagger }\hat{a}+\frac{1}{2}\hbar \omega 
\hat{\sigma}_{z},  \label{H_single1} \\
\hat{H_{1}} &=&\hbar g(\hat{\sigma}_{_{+}}\hat{a}+\hat{\sigma}_{_{-}}\hat{a}%
^{\dagger }).  \label{H_single2}
\end{eqnarray}%
The operator $\hat{\sigma}_{_{+}}$ leads the atom from state $|b\rangle $ to
state $|a\rangle $, whereas $\hat{\sigma}_{_{-}}$ makes the inverse
operation. We should also define 
\begin{eqnarray}
\hat{\sigma}_{z} &=&|a\rangle \langle a|-|b\rangle \langle b|,\hat{\sigma}%
_{x}=(\hat{\sigma}_{_{+}}+\hat{\sigma}_{_{-}}),  \label{spin} \\
\hat{\sigma}_{_{y}} &=&-i(\hat{\sigma}_{_{+}}-\hat{\sigma}_{_{-}}), \label{spin_y}
\end{eqnarray}%
operators which fulfill the Pauli algebra ($\left[ \hat{\sigma}_{i},\hat{%
\sigma}_{j}\right] =2i\epsilon _{ijk}\hat{\sigma}_{k}).$

Considering the case the electric field is linearly polarized in the
x-direction and $\mathbf{P}_{ab}$ is real \cite{foot2}, we can write 
\begin{equation}
g=-\frac{e{P}_{ab}E_{0}}{\hbar }.  \label{g}
\end{equation}%
The normalization factor is equal to 
\begin{equation}
E_{0}=\sqrt{\hbar \nu /(2\epsilon _{0}\mathcal{V)}}.  \label{eqE0}
\end{equation}%
Experimentally the cavity set-up provides a precise description for the
atomic dynamics even with the atom-field Hamiltonian $\hat{H_{1}}$ given by
Eq. (\ref{H_single2}), since the interaction with a single mode dominates
the evolution.

In the interaction picture, the interaction potential $\hat{V}$ is defined
as 
\begin{equation}
\hat{V}=\hat{U}_{0}^{\dagger }(t)\hat{H}_{1S}\hat{U}_{0}(t),  \label{V}
\end{equation}%
with $\hat{H}_{1S}$ being the Schr\"{o}dinger (time independent)
representation of $\hat{H}_{1S}$, and $\hat{U}_{0}(t)=e^{-i\hat{H_{0}}%
t/\hbar }$ is the time evolution operator in this picture. In order to
evaluate $\hat{V},$ the Baker-Campbell-Hausdorff lemma is applied and
yields: 
\begin{eqnarray}
e^{i\nu \hat{a}^{\dagger }\hat{a}t}\hat{a}e^{-i\nu \hat{a}^{\dagger }\hat{a}%
t} &=&\hat{a}e^{-i\nu t},  \label{BCHa} \\
e^{i\nu \hat{a}^{\dagger }\hat{a}t}\hat{a}^{\dagger }e^{-i\nu \hat{a}%
^{\dagger }\hat{a}t} &=&\hat{a}^{\dagger }e^{i\nu t},  \label{BCHadag} \\
e^{\frac{1}{2}i\omega \hat{\sigma}_{z}t}\hat{\sigma}_{\pm }e^{-\frac{1}{2}%
i\omega \hat{\sigma}_{z}t} &=&\hat{\sigma}_{_{\pm }}e^{\pm i\omega t}.
\label{b}
\end{eqnarray}%
Replacing the former relations and Eq. (\ref{H_single2}) on Eq. (\ref{V}),
we get 
\begin{equation}
\hat{V}=\hbar g(\hat{\sigma}_{_{+}}\hat{a}e^{i\Delta t}+\hat{a}^{\dagger }%
\hat{\sigma}_{_{-}}e^{-i\Delta t}),
\end{equation}%
with $\Delta =\left( \omega -\nu \right) $. In this representation, the
wavefunction $|\psi _{I}(t)\rangle =\widehat{U}_{0}^{\dagger }(t)|\psi
_{S}(t)\rangle $ is a linear combination of the two atomic states with
arbitrary number of photons $(n)$ in the cavity ($|a,n\rangle $ and $%
|b,n\rangle )$. So we have 
\begin{equation}
|\psi _{I}(t)\rangle =\sum_{n=0}^{\infty }[c_{a,n}|a,n\rangle
+c_{b,n}|b,n\rangle ].
\end{equation}%
The Schr\"{o}dinger equation in the interaction picture $i\hbar |\dot{\psi}%
_{I}(t)\rangle =\hat{V}|\psi _{I}(t)\rangle $ leads to the following system
of coupled differential equations: 
\begin{eqnarray}
\dot{c}_{a,n} &=&-ig\sqrt{n+1}e^{i\Delta t}c_{b,n+1}, \\
\dot{c}_{b,n+1} &=&-ig\sqrt{n+1}e^{-i\Delta t}c_{a,n},
\end{eqnarray}%
which can be easily solved. Considering the atom initially in the excited
state $|a\rangle $ and the field with a distribution $c_{n}(0)$ of photons,
we can write $c_{a,n}(0)=c_{n}(0)$ and $c_{b,n+1}(0)=0$ to get (see Ref. 
\cite{scully}, chap. 6): 
\begin{eqnarray}
c_{a,n}(t) &=&c_{n}(0)e^{i\Delta t/2}[\cos (\gamma t)-\frac{i\Delta }{\Omega
_{n}}\sin (\gamma t)], \\
c_{b,n+1}(t) &=&-c_{n}(0)\frac{2ig\sqrt{n+1}}{\Omega _{n}}\sin (\gamma
t)e^{-i\Delta t/2},
\end{eqnarray}%
where $\Omega _{n}^{2}=\Delta ^{2}+4g^{2}(n+1)$ and $\gamma =\Omega _{n}/2$.
The atomic inversion function can be now defined as 
\begin{equation}
W(t)=\sum_{n=0}^{\infty }(|c_{a,n}(t)|^{2}-|c_{b,n}(t)|^{2}),
\end{equation}%
Here, the discrete character of the sum over the number of photons is crucial
for the observation of periodic collapses and revivals of the inversion
function as a pure quantum effect (see Ref.\cite{walls}, chap. 10). For the
particular case of resonance ($\Delta =0$) we have 
\begin{equation}
W(t)=\sum_{n=0}^{\infty }\rho _{nn}(0)\cos (2\sqrt{n+1}gt),  \label{W_coh}
\end{equation}%
where $\rho _{nn}(0)=|c_{n}(0)|^{2}$ is the probability that there are $n$
photons in the cavity at $t=0$.

Considering an initial coherent state in a cavity with medium number of
photons $\bar{n}$ we obtain 
\begin{equation}
W(t)=\sum_{n=0}^{\infty }\frac{\bar{n}^{n}e^{-\bar{n}}}{n!}\cos \text{(}2%
\sqrt{n+1}gt\text{)}.
\end{equation}%
For $\bar{n}>>1$ (but not so large, in order to fulfill the RWA
condition $g\sqrt{\bar{n}}\ll \omega $ \cite{rwa}), with small variance $%
\Delta n,$ we recover the known Rabi nutation with frequency $\Omega _{\bar{n%
}}^{0}\simeq 2g\sqrt{\bar{n}}$. This is the semiclassical expression as
expected from the correspondence principle. For intermediate values of $\bar{%
n}$, this expression leads to the characteristic behavior of collapses and
revivals of the population inversion. This known effect appears mainly due
to the interference of the several oscillating patterns associated to
different photon numbers.

In particular, at resonance and taking the vacuum ($\rho _{nn}(0)=\delta
_{n,0}$) as the initial state in the cavity, we find $W(t)=\cos (2gt)$. So,
Rabi oscillations with frequency $\Omega =2g$ occur due to spontaneous
emission, a typical quantum effect that does not occur in the semi-classical
system.

%%%%%%%%%%%%%%%%%%%%%%%%%%%%%%%%%%%%%%%%%%%%%%%%%%%%%%%%%%%%%%%%%%%%%%%%%%

\section{Lorentz-violation effects}

Now we will explore how these quantum fundamental effects can be affected
depending on the strength of the Lorentz-breaking coupling in Lagrangian (%
\ref{L1}). The Hamiltonian can now be written as 
\begin{equation}
\hat{H}=\hat{H_{0}}+\hat{H_{1}}+\hat{H_{1}^{\prime }}+\hat{H_{2}^{\prime }},
\end{equation}%
where now we have the additional LV contributions: $\hat{H_{1}^{\prime }}=e%
\mathbf{A}\cdot \mathbf{v}/m_{e}$ and $\hat{H_{2}^{\prime }}=-\mathbf{p}%
\cdot \mathbf{v}/m_{e}$.

First of all, we analyze the $\hat{H_{1}^{\prime }}=e\widehat{A}_{x}$v$%
_{x}/m_{e}$ contribution. After quantization, in the time-independent Schr%
\"{o}dinger representation, we can write 
\begin{equation}
\hat{H_{1}^{\prime }}=\frac{ieE_{0}\text{v}_{x}}{m_{e}\nu }(-\hat{a}+\hat{a}%
^{\dagger }).  \label{eqH1pr}
\end{equation}%
In order to obtain the interaction potential $\hat{V}_{1}^{\prime }=\hat{U}%
_{0}^{\dagger }(t)\hat{H}_{1}^{\prime }\hat{U}_{0}(t)$, we use Eqs. (\ref%
{eqH1pr}), (\ref{BCHa}) and (\ref{BCHadag}), leading to 
\begin{equation}
\hat{V}_{1}^{\prime }=\frac{ieE_{0}\text{v}_{x}}{m_{e}\nu }(-\hat{a}e^{-i\nu
t}+\hat{a}^{\dagger }e^{i\nu t}).
\end{equation}

Now we turn to the contribution of $\hat{H_{2}^{\prime }}=-\widehat{p}_{x}$v$%
_{x}/m_{e}=-\widehat{\dot{x}}$v$_{x}$. The $\widehat{\dot{x}}$ operator can
be rewritten using the Heisenberg equation in the interaction picture as ${%
\widehat{H}}_{2}^{\prime }=-$v$_{x}/(i\hbar )[\widehat{x},\widehat{H}_{0}]$.
We found better to represent the $\hat{x}$ operator as 
\begin{equation}
\hat{x}=-iP_{ab}\hat{\sigma}_{y}\hat{\sigma}_{z},  \label{x_op}
\end{equation}%
with $P_{ab}\equiv \langle a|\hat{x}|b\rangle =P_{ab}^{\ast }=P_{ba}$, where
the operators $\hat{\sigma}_{z},$ $\hat{\sigma}_{y}$ are defined in Eqs. (\ref%
{spin})-(\ref{spin_y}). If we represent the energy eigenstates in a vector form 
\begin{equation}
|a\rangle =\left( 
\begin{array}{c}
1 \\ 
0%
\end{array}%
\right) ,\,\,\,|b\rangle =\left( 
\begin{array}{c}
0 \\ 
1%
\end{array}%
\right) ,
\end{equation}%
we can identify the operators $\hat{\sigma}_{x},\hat{\sigma}_{y},\hat{\sigma}%
_{z}$ with the Pauli matrices 
\begin{equation}
\hat{\sigma}_{x}=\left( 
\begin{array}{cc}
0 & 1 \\ 
1 & 0%
\end{array}%
\right) ,\,\,\,\hat{\sigma}_{y}=\left( 
\begin{array}{cc}
0 & -i \\ 
i & 0%
\end{array}%
\right) ,\,\,\,\hat{\sigma}_{z}=\left( 
\begin{array}{cc}
1 & 0 \\ 
0 & -1%
\end{array}%
\right) .
\end{equation}%
Some remarks about the mathematical notation are worthy. Pauli matrix $%
\sigma _{z}$ was already used in Eq. (\ref{H_single1}) for the free
Hamiltonian as an economic way to describe the atom energy content. Hence,
the use of Pauli matrices in this way has no connection with spin operators
or spin states. These so-called pseudo-spin operators (Ref. \cite{walls}, p.
203) are very useful to simplify the calculations and should not lead us to
misunderstandings concerning spin magnetic features of the atom states.

In order to achieve a description in the interaction picture, we must apply
again the Baker-Campbell-Hausdorff lemma to obtain 
\begin{equation}
e^{\frac{1}{2}i\omega \hat{\sigma}_{z}t}\hat{x}e^{-\frac{1}{2}i\omega \hat{%
\sigma}_{z}t}=-iP_{ab}[\sin (\omega t)\hat{\sigma}_{x}+\cos (\omega t)\hat{%
\sigma}_{y}]\hat{\sigma}_{z},
\end{equation}%
where Eq. (\ref{x_op}) was used. In this representation, the
interaction potential 
\begin{equation}
\hat{V}_{2}^{\prime }=-\frac{\text{v}_{x}}{i\hbar }[e^{i\hat{H}_{0}t/\hbar }%
\hat{x}e^{-i\hat{H}_{0}t/\hbar }\hat{H}_{0}-\hat{H}_{0}e^{i\hat{H}%
_{0}t/\hbar }\hat{x}e^{-i\hat{H}_{0}t/\hbar }],
\end{equation}%
takes the form: 
\begin{equation}
\hat{V}_{2}^{\prime }=i\text{v}_{x}P_{ab}\omega \lbrack \cos (\omega t)\hat{%
\sigma}_{x}-\sin (\omega t)\hat{\sigma}_{y}]\hat{\sigma}_{z}.
\end{equation}

The Schr\"{o}dinger equation in the interaction picture, $i\hbar |\dot{\psi}%
_{I}(t)\rangle =(\hat{V}+\hat{V}_{1}^{\prime }+\hat{V}_{2}^{\prime })|\psi
_{I}(t)\rangle $, yields a system of coupled differential equations for the
probability amplitudes: 
\begin{eqnarray}
\dot{c}_{a,n} &=&-ig\sqrt{n+1}e^{i\Delta t}c_{b,n+1}  \notag  \label{eom1} \\
&&+\frac{eE_{0}\text{v}_{x}}{m_{e}\hbar \nu }(-c_{a,n+1}\sqrt{n+1}e^{-i\nu
t}+c_{a,n-1}\sqrt{n}e^{i\nu t})  \notag \\
&&-\frac{\text{v}_{x}P_{ab}}{\hbar }\omega e^{i\omega t}c_{b,n}, \\
\dot{c}_{b,n+1} &=&-ig\sqrt{n+1}e^{-i\Delta t}c_{a,n}  \notag  \label{eom2}
\\
&&+\frac{eE_{0}\text{v}_{x}}{m_{e}\hbar \nu }(-c_{b,n+2}\sqrt{n+2}e^{-i\nu
t}+c_{b,n}\sqrt{n+1}e^{i\nu t})  \notag \\
&&+\frac{\text{v}_{x}P_{ab}}{\hbar }\omega e^{-i\omega t}c_{a,n+1}.
\end{eqnarray}%
with $n=0,1,...\infty $. In general, this system of infinite equations is of
difficult solution. Here, we will study the system at resonance ($\Delta =0$%
).

For large $n$ we have 
\begin{eqnarray}
\dot{c}_{a,n} &=&-ig\sqrt{n}c_{b,n}  \notag \\
&&+\frac{eE_{0}\text{v}_{x}}{m_{e}\hbar \nu }c_{a,n}\sqrt{n}2i\sin (\nu t) 
\notag \\
&&-\frac{\text{v}_{x}P_{ab}}{\hbar }\omega e^{i\omega t}c_{b,n}, \\
\dot{c}_{b,n} &=&-ig\sqrt{n}c_{a,n}  \notag \\
&&+\frac{eE_{0}\text{v}_{x}}{m_{e}\hbar \nu }c_{b,n}\sqrt{n}2i\sin (\nu t) 
\notag \\
&&+\frac{\text{v}_{x}P_{ab}}{\hbar }\omega e^{-i\omega t}c_{a,n}.
\end{eqnarray}%
This must be compared with the differential equations for the coefficients
obtained in the semiclassical theory\cite{Semi}: 
\begin{align}
\overset{\cdot }{a}(t)& =i(\Omega _{R}/2)b(t)+i\alpha _{0}a(t)\sin \nu
t-\beta _{0}\omega b(t)e^{i\omega t},  \label{a2} \\
\overset{\cdot }{b}(t)& =i(\Omega _{R}/2)a(t)+i\alpha _{0}b(t)\sin \nu
t+\beta _{0}\omega a(t)e^{-i\omega t},  \label{b2}
\end{align}%
where $\alpha _{0}=eE_{0}^{sc}$v$_{x}/(m_{e}\hbar \nu )$, $\beta _{0}=($v$%
_{x}P_{ab}/hbar)$, $\Omega_R$ is the Rabi frequency and $E_{0}^{sc}=\sqrt{2n\hbar \nu /\epsilon _{0}\mathcal{V}%
}$ is the semiclassical expression corresponding to the normalization factor 
$E_{0}$. Note that $2\sqrt{n}E_{0}=E_{0}^{sc}$ and the correspondence
principle is verified.

%%%%%%%%%%%%%%%%%%%%%%%%%%%%%%%%%%%%%%%%%%%%%%%%%%%%%%%%%%%%%%%%%%%%%%%%%%%%%%%%%%%%

We consider circular Rydberg atoms in a high Q cavity. In such atoms the
long radiative lifetime makes atomic relaxation negligible during the atom's
transit time across the cavity \cite{prl96}. Also a high Q cavity turns the
photon lifetime longer than the atom-cavity interaction time. Brune et al. 
\cite{prl94} investigated the resonant effects of the vacuum in a cavity
mode. There the authors considered the transition frequency $\omega =51GHz$
for $n=50$ circular Rydberg atoms. The matrix element between the circular
levels 50 and 51 of a linear projection of the electric dipole on the
orbit's plane has the large value $eP_{ab}=1250ea_{0}$, where $a_{0}$ is the
Bohr radius. This means a very large classical radius and a very large
radiative decay time.  The effective cavity volume was $0.7cm^{3}$. With it,
Eq. (\ref{eqE0}) provides a vacuum field amplitude at antinodes of $%
E_{0}=6.95\times 10^{-4}V/m$. These parameters correspond to the following
vacuum Rabi oscillation frequency $\Omega _{vac}=2g={2eP_{ab}E_{0}}/\hbar
=132kHz$, which implies $g=66kHz$ for the coupling constant. In these
estimates we are neglecting the usual spatial variation of the
electromagnetic field in the cavity. We will also consider that the applied
field frequency is near resonance $\left( \nu =\omega =51GHz\right) $.

Now, we can use the above values for the parameters to estimate the relative
importance of the Lorentz-violating terms in Eqs. (\ref{eom1})-(\ref{eom2}).
In this way, we have $eE_{0}/(m_{e}\hbar \omega )=2\times
10^{31}kg^{-1}m^{-1}$ and $\ P_{ab}\omega /{\hbar }=3\times
10^{37}kg^{-1}m^{-1}$. Then, we note that for $\bar{n}\ll 10^{12}$ the
magnitude of the second term (corresponding to the influence of $-\mathbf{p}%
\cdot \mathbf{v}/m_{e}$) can be taken as much larger than the magnitude of
the first one (corresponding to $e\mathbf{A}\cdot \mathbf{v}/m_{e}$). At
resonance, both terms oscillate with the same frequency, but the smaller
amplitude argument is an enough reason to neglect the term stemming from $e%
\mathbf{A}\cdot \mathbf{v}/m_{e}$. The fact that such term does not imply
quantum effects at a first approximation is in agreement with the
semi-classical behavior \cite{Semi} associated with it. Indeed, it amounts
only at phase effects without altering the semi-classical population
inversion function. Neglecting this term, it is ascribed an explicit
gauge-independent character for the LV modifications.

From these considerations the equations for the probability amplitudes can
be properly approximated as

\begin{eqnarray}
\dot{c}_{a,n} &\simeq &-ig\sqrt{n+1}c_{b,n+1}  \notag  \label{eom3} \\
&&-\frac{\text{v}_{x}P_{ab}}{\hbar }\omega e^{i\omega t}c_{b,n}, \\
\dot{c}_{b,n+1} &\simeq &-ig\sqrt{n+1}c_{a,n}  \notag  \label{eom4} \\
&&+\frac{\text{v}_{x}P_{ab}}{\hbar }\omega e^{-i\omega t}c_{a,n+1}.
\end{eqnarray}

Here, an interesting issue is to know if the high frequency of the
Lorentz-violating term can in fact provide an effective correction. In fact,
we remind that a high-frequency term proportional to $2\omega $ was
discarded due to the rotating-wave approximation. The exclusion of the
counter-rotating terms from the equations of motion is justified when the
frequency of the external modes most strongly interacting with the system is
very large compared to the strength $g$ of the interaction \cite{walls} (in
the present case $\omega =51GHz$ whereas $g=66kHz$). We have already seen
that such exclusion is equivalent to neglecting the non-energy-conserving
processes such as the excitation of an atom along with the emission of a
photon \cite{walls}. From this point of view the Lorentz-violating term
induced by the background has an energy non-conserving character, once the
Hamiltonian of Eq. (\ref{eqH1pr}) amounts at creation and annihilation of a
photon without a corresponding change on the atomic level. One possibility
is to consider physical situations where such an energy-violating terms
grows in importance (see \cite{mil94}, p. 151) and \cite{cook}. However, as
the terms depending on the background oscillate with frequency $\omega $
whereas the discarded terms from the RWA oscillate with $2\omega $, the
investigation of the influence of the background on the modified equations
of motion (Eqs. (\ref{eom3})-(\ref{eom4})) seems to be a sensible option.

We can decouple those equations using the approximation that during the time
interval of measurement of Rabi nutation the oscillating terms are averaged
to zero [$\left\langle \cos (\omega t)\right\rangle =\left\langle \sin
(\omega t)\right\rangle =0]$. This gives 
\begin{eqnarray}
\ddot{c}_{a,n} &\simeq &\biggl(-g^{2}{(n+1)}-\frac{\text{v}%
_{x}^{2}P_{ab}^{2}\omega ^{2}}{\hbar ^{2}}\biggr)c_{a,n}\,, \\
\ddot{c}_{b,n+1} &\simeq &\biggl(-g^{2}{(n+1)}-\frac{\text{v}%
_{x}^{2}P_{ab}^{2}\omega ^{2}}{\hbar ^{2}}\biggr)c_{b,n+1}\,.
\end{eqnarray}%
We consider as the initial state the atom in the excited state with the
cavity with a field characterized by coefficients $c_{n}(0)$, so that the
results are 
\begin{eqnarray}
c_{a,n} &\simeq &c_{a,n}(0)\cos \biggl[\zeta _{n}\sqrt{n+1}gt\biggr]\,,
\label{CAn_sol} \\
c_{b,n+1} &\simeq &-ic_{a,n}(0)\sin \biggl[\zeta _{n}\sqrt{n+1}gt\biggr]\,.
\label{CBn_sol}
\end{eqnarray}%
with

\begin{eqnarray}
\zeta_n\equiv\sqrt{1+\frac1{n+1}\biggl(\frac{ \text{v}_{x}P_{ab}\omega}{%
\hbar g}\biggr)^2}
\end{eqnarray}

The average number of photons is 
\begin{equation*}
\bar{n}(t)=\sum_{0}^{N}(nP_{n}(t)).
\end{equation*}%
where $P_{n}(t)=|c_{a,n}|^{2}+|c_{b,n}|^{2}$ is the probability for finding $%
n$ photons in the cavity. Now note from Eqs. (\ref{CAn_sol})-(\ref{CBn_sol})
that $P_{n}(t)=P_{n}(0)$. This shows that the photon statistics and
consequently the average number of photons is not altered by the background v%
$_{x}$.

Considering an initial vacuum state in a cavity at resonance ($\Delta =0$),
we attain the following population inversion function: 
\begin{equation}
W(t)=\cos (2g\zeta _{0}t).  \label{W_vac}
\end{equation}%
For a sufficiently small background $\left( {\text{v}_{x}P_{ab}\nu /\hbar
g\ll 1}\right) $, it is allowed to write 
\begin{equation}
\zeta _{0}\cong 1+\frac{1}{2}\frac{\text{v}_{x}^{2}P_{ab}^{2}\nu ^{2}}{\hbar
^{2}g^{2}}  \label{zeta_vac}
\end{equation}%
The former Eqs.(\ref{W_vac})-(\ref{zeta_vac}) mean that, at first
approximation, there appears an effective coupling (due to the background)
given by $g^{\prime }=g+{\text{v}_{x}^{2}P_{ab}^{2}\nu ^{2}}/({2\hbar ^{2}g}%
) $. Consequently, this implies an increasing on the value of Rabi
frequency: 
\begin{equation}  \label{Omega_0}
\Omega _{0}=2g\biggl(1+\frac{1}{2}\frac{\text{v}_{x}^{2}P_{ab}^{2}\nu ^{2}}{%
\hbar ^{2}g^{2}}\biggr)
\end{equation}%
Regarding that discrepancies of 1 part in $10^{10}$ from the usual results
of quantum mechanics can be detected, we can limit the LV effects in
accordance with this sensitivity. We then impose that the correction term
should be smaller than $10^{-10}$, that is (v$_{x}P_{ab}\nu /\hbar
g)<10^{-10}.$ For the chosen parameters\textbf{\ }$P_{ab}\omega /\hbar
=3.2\times 10^{37}kg^{-1}m^{-1}$ and $g=66kHz$,\textbf{\ s}uch condition
leads to the following upper bound on the LV background: \textbf{$\text{v}$}$%
_{x}<2.06\times 10^{-38}kgm/s$, or in natural units $\text{v}$$%
_{x}<10^{-10}eV$\textbf{. }

The same effect can be studied for an initial coherent state in a cavity
with medium number of photons $\bar{n}$ and at resonance ($\Delta =0$). In
this case we attain the following population inversion function: 
\begin{equation}
W(t)=\sum_{n=0}^{\infty }\frac{{\bar{n}}^{n}e^{-{\bar{n}}}}{n!}\cos \biggl[2%
\sqrt{n+1}g\sqrt{1+\frac{\text{v}_{x}^{2}P_{ab}^{2}\omega ^{2}}{(n+1)\hbar
^{2}g^{2}}}t\biggr].  \label{Wt_lv}
\end{equation}%
This expression allows to infer that the net effect of the background on the
probability amplitudes is the increasing of the frequency of the collapses
and revivals of the population inversion. It is instructive to note how the
high frequency terms of opposite phases in Eqs. (\ref{eom3})-(\ref{eom4})
conspired to provide such a correction. Note also that for v$_{x}=0$ the
usual result for the inversion $W(t)$ - a superposition of frequencies $2%
\sqrt{n+1}g$ - is recovered for an initial coherent state (see Eq. (\ref%
{W_coh})). For large $\bar{n},$ the sum in Eq. (\ref{Wt_lv}) can be
simplified to produce $W(t)\sim \cos (\Omega _{\bar{n}}t)$, where 
\begin{equation}
\Omega _{\bar{n}}=2\sqrt{\bar{n}}g\sqrt{1+\frac{\text{v}_{x}^{2}P_{ab}^{2}%
\omega ^{2}}{\bar{n}\hbar ^{2}g^{2}}}
\end{equation}%
is the Rabi frequency corrected by the Lorentz-violating background. This
expression reveals that the Rabi frequency increases with the background
magnitude. For small background we can also write 
\begin{equation}
\Omega _{\bar{n}}\cong 2\sqrt{\bar{n}}g\biggl({1+\frac{1}{2}\frac{\text{v}%
_{x}^{2}P_{ab}^{2}\omega ^{2}}{\bar{n}\hbar ^{2}g^{2}}}\biggr)  \label{O_lv}
\end{equation}

An important point here is the appearance of the mean photon number factor ($%
\bar{n})$ in the correction term in comparison with the previous case of an
initial vacuum state in the cavity (see Eq. (\ref{Omega_0})). Indeed, the
larger the average number of photons for an initial coherent state in the
cavity, the lower is the correction the Rabi frequency induced by the
fixed background. Considering the same experimental sensitivity of 1 part in 
$10^{10}$ in a measurement of the Rabi frequency, a larger bound (less
stringent) on $\text{v}_{x}$ may now be achieved. In fact, for a state with
large number of photons $\left( \bar{n}>>1\right) $, the allowed background
upper value is multiplied by $\sqrt{\bar{n}}$ ($\text{v}_{x}^{COH}=\sqrt{%
\bar{n}}\text{v}_{x}^{VAC}$\textbf{$).$} This means that cavity experiments
with smaller number of photons imply better bounds (more stringent) on the
background magnitude. In this sense, the best probes for determining LV\
deviations from usual quantum mechanics are really the vacuum states.

%%%%%%%%%%%%%%%%%%%%%%%%%%%%%%%%%%%%%%%%%%%%%%%%%%%%%%%%%%%%%%%%%%%%%%%%%%%%%%%%%%%%%%%%%%%%%%%%%%%%%%%%%%%%%%%%%%%%
\begin{figure}[th]
\begin{center}
\includegraphics[{angle=0,width=8.5cm}]{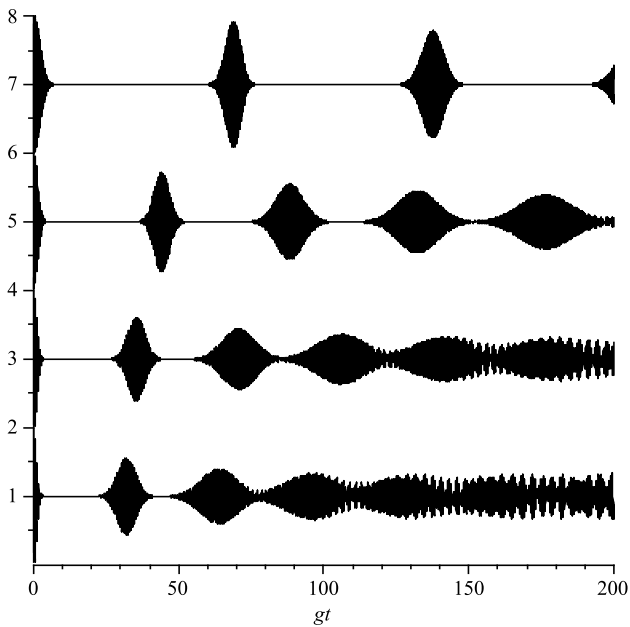}
\end{center}
\caption{Population inversion as a function of time $C+W(t)$, for $\bar{n}%
=25 $ and (a) $C=1,\protect\nu _{x}=0$ (first, lower curve), (b) $C=3,%
\protect\nu _{x}=5\times10^{-33}kgm/s$ ( $2.5\times 10^{-5}eV$ in natural
units - second curve), (c) $C=5,\protect\nu _{x}=1\times 10^{-32}kgm/s$ ( $%
5\times10^{-5}eV$ in natural units - third curve) and (d) $C=7,\protect\nu %
_{x}=2\times10^{-32}kgm/s$ (or $\protect\nu _{x}=1\times 10^{-4}eV$ -
fourth, upper curve). The $C$ constants are included to make easier the
comparison.}
\label{fig_Wt}
\end{figure}
%%%%%%%%%%%%%%%%%%%%%%%%%%%%%%%%%%%%%%%%%%%%%%%%%%%%%%%%%%%%%%%%%%%%%%%%%%%%%%%%%%%%%%%%%%%%%%%%%%%%%%%%%%%%%%%%%%%%  
To study the intermediate scale where effects of collapses and revivals
appear and to see the influence of the extra factor depending on the
background v$_{x}$ on the expression of Eq. (\ref{Wt_lv}), we proceed with a
graphical analysis. In Fig. 1, we present a plot of $W(t)\times gt$. Such a
figure shows a sequence of four curves for an initial coherent state with $%
\bar{n}=25$ and coupling $g=66kHz$. The first lower curve shows the usual
sequence of collapses and revivals in the absence of Lorentz violation $%
\left( \text{v}_{x}=0\right) $. As the background increases, the collapses
and revivals tend to occur later. Note also that the sequence of collapses
and revivals is always destroyed after some time, larger for higher values
of the background. These observations indicate that a stronger background
favours the maintenance of the revival/collapse sequence for a greater time.

%%%%%%%%%%%%%%%%%%%%%%%%%%%%%%%%%%%%%%%%%%%%%%%%%%%%%%%%%%%%%%%%%%%%%%%%%%%%%%%%%%%%%%%%%%%%%%%%%%%%%%%%%%%%%%%%%%%%
\begin{figure}[th]
\begin{center}
\includegraphics[{angle=0,width=8.5cm}]{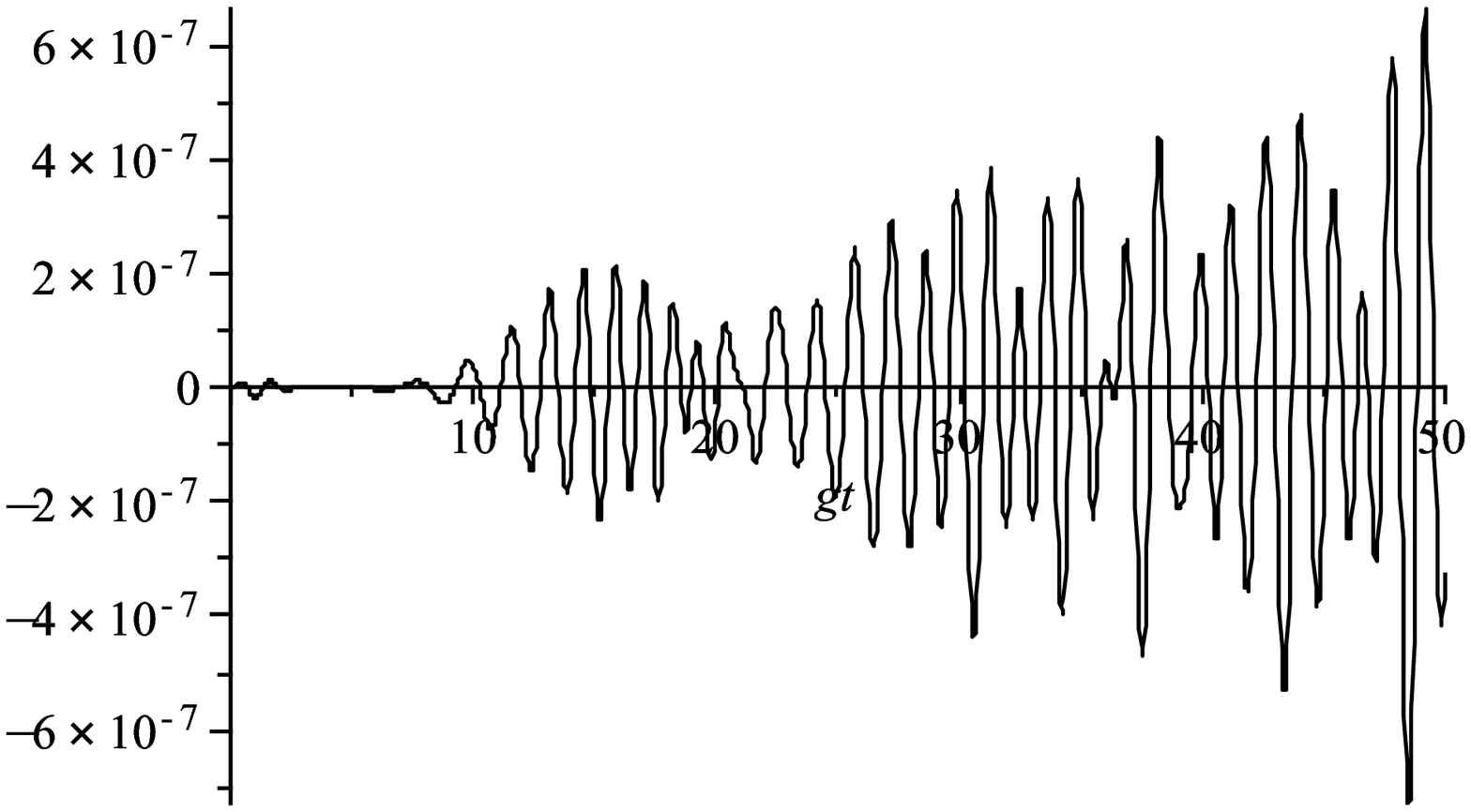} %
\includegraphics[{angle=0,width=8.5cm}]{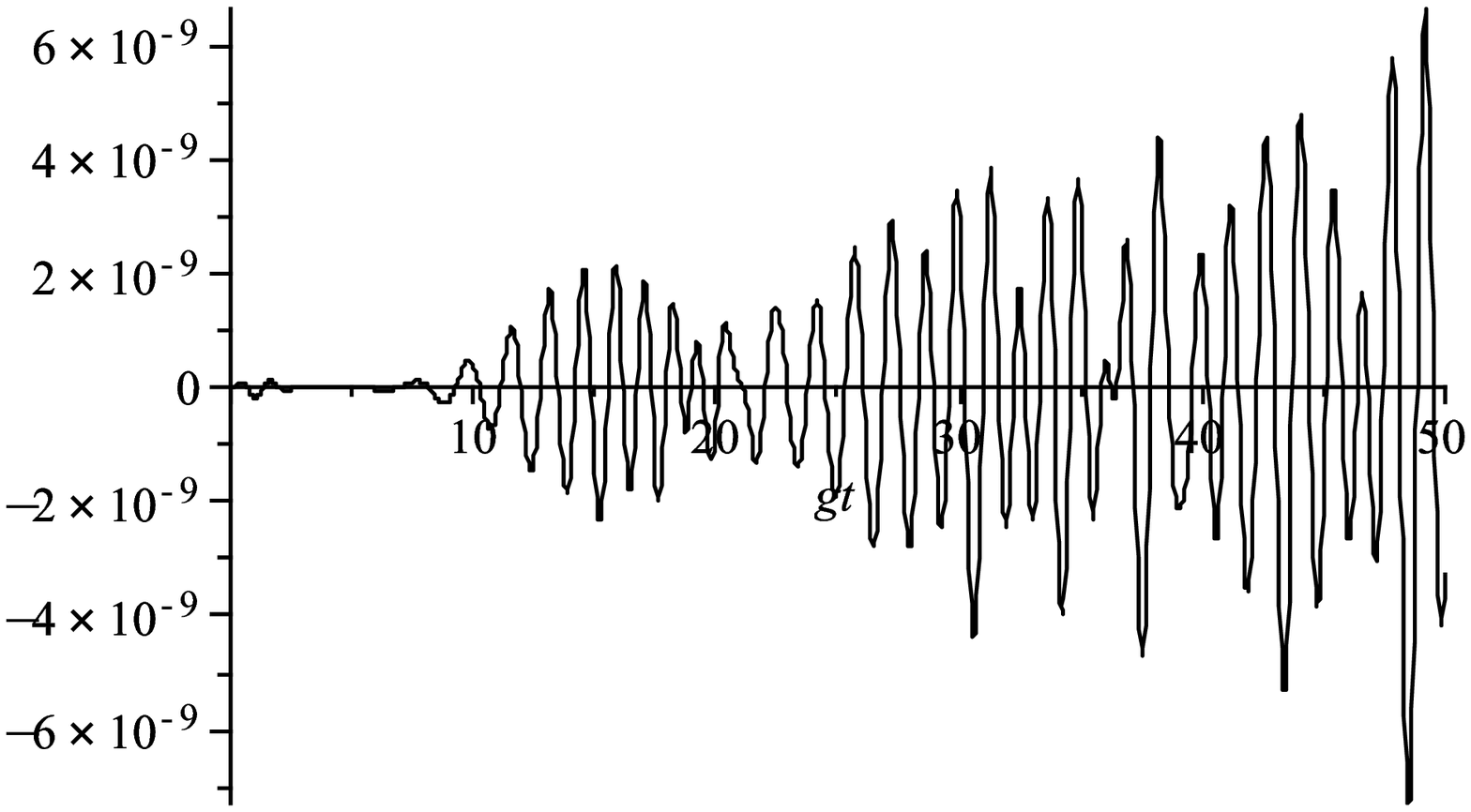} %
\includegraphics[{angle=0,width=8.5cm}]{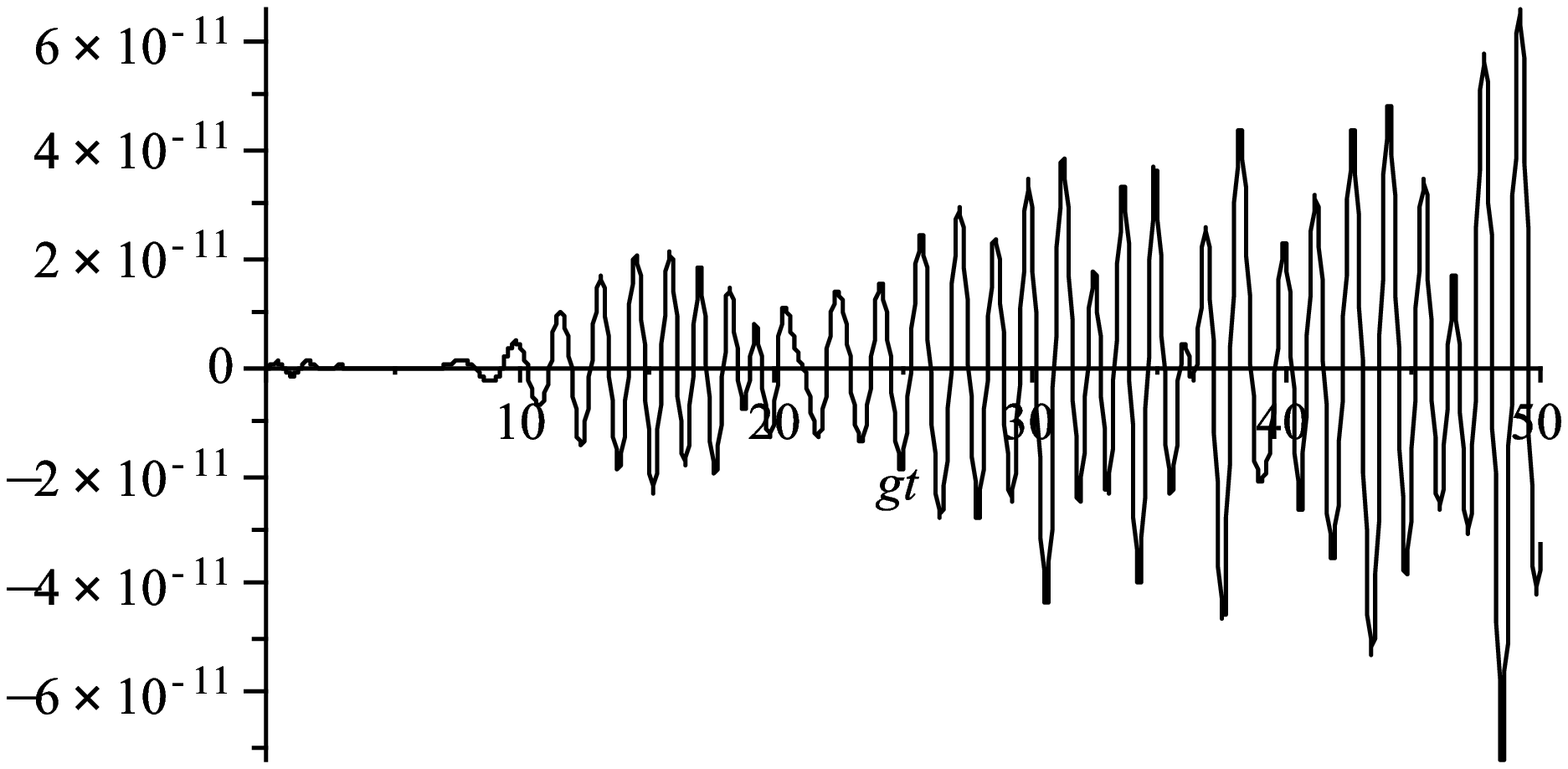}
\end{center}
\caption{Difference $W-W_0$ between population inversion functions $W$ for $%
\text{v}_x$ and $W_0$ for $\text{v}_x=0$, for $\bar{n}=25 $ where (a) $%
\protect\nu _{x}=5\times10^{-37}kgm/s$ ( $2.5\times 10^{-9}eV$ in natural
units - upper curve), (b) $\protect\nu _{x}=5\times10^{-38}kgm/s$ ( $%
2.5\times 10^{-10}eV$ in natural units - middle curve), (c) $\protect\nu %
_{x}=5\times10^{-39}kgm/s$ ( $2.5\times 10^{-11}eV$ in natural units - lower
curve)}
\label{fig_Wt}
\end{figure}
%%%%%%%%%%%%%%%%%%%%%%%%%%%%%%%%%%%%%%%%%%%%%%%%%%%%%%%%%%%%%%%%%%%%%%%%%%%%%%%%%%%%%%%%%%%%%%%%%%%%%%%%%%%%%%%%%%%%  

An interesting issue is to verify if this kind of analysis is able to impose
an upper bound on the background magnitude, in agreement to the analytical
result of Eq. (\ref{O_lv}). A reasonable criterion consists in taking the
maximum background value that does not yield significative discrepancy (of 1
part in $10^{10}$) on the population inversion pattern taking as reference
the usual case (v$_{x}=0)$. In this way, we have chosen to plot in Fig. 2
the difference $W-W_{0}$ at the time scale $0<gt<50$, sufficient to reveal
the beginning of the second collapse in the usual case (see lower picture in
Fig. 1). The comparative results from Fig. 2 shows that $\text{v}%
_{x}<10^{-10}eV$ is an efficient condition in keeping the difference $W-W_{0}
$ below $10^{-10}$ for the first observed sequence of collapses and revivals
of the population inversion. This is in agreement with our previous estimate
for the bound for an initial vacuum state in the cavity.

Fig. 3 shows the same kind of graphical analysis of Fig. 1 for an initial
coherent state with lower mean number of photons $\left( \bar{n}=5\right) $.
Here, the background effect of keeping the sequence of collapses and
revivals for longer times is more clearly depicted. Further, it is seen that
the collapse time and separation time (between the revivals) increase with
the background magnitude. However, even lower background
values already reveal larger separations between collapses and revivals (in
comparison with the case of Fig. 1).

%%%%%%%%%%%%%%%%%%%%%%%%%%%%%%%%%%%%%%%%%%%%%%%%%%%%%%%%%%%%%%%%%%%%%%%%%%%%%%%%%%%%%%%%%%%%%%%%%%%%%%%%%%%%%%%%%%%%
\begin{figure}[th]
\begin{center}
\includegraphics[{angle=0,width=8.5cm}]{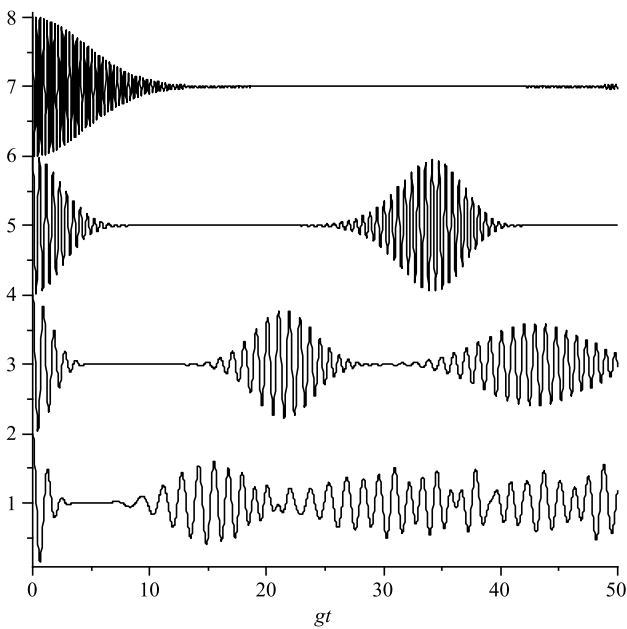}
\end{center}
\caption{Population inversion as a function of time $C+W(t)$, for $\bar{n}=5$%
. Values for $C$ and ${\text v}_x$ are the same from Fig. 1.}
\label{fig_Wt2}
\end{figure}
%%%%%%%%%%%%%%%%%%%%%%%%%%%%%%%%%%%%%%%%%%%%%%%%%%%%%%%%%%%%%%%%%%%%%%%%%%%%%%%%%%%%%%%%%%%%%%%%%%%%%%%%%%%%%%%%%%%%  

A further step is to use Eq. (\ref{Wt_lv}) to study the implied changes on
the known expressions for the times $t_{R}$ (period of Rabi oscillations), $%
t_{c}$ (collapse time) and $t_{r}$ (revival time), defined in Ref. \cite%
{scully}. As before, we consider $\bar{n}\gg 1$. To measure the influence of
the background, it will be used the dimensionless parameter $\alpha =$v$%
_{x}P_{ab}\omega /(\hbar g)$. The time period $t_{R}$ is given by the
inverse of Rabi frequency%
\begin{equation}
t_{R}\sim \frac{1}{\Omega _{\bar{n}}}=\frac{1}{\Omega _{\bar{n}}^{0}}\bigg(1-%
\frac{\alpha ^{2}}{\bar{n}}\biggr),  \label{eqtR}
\end{equation}%
outcome obtained using Eq. (\ref{O_lv}), with $\Omega _{\bar{n}}^{0}=2g\sqrt{%
\bar{n}}$. It shows that $t_{R}$ is reduced by the presence of the
background. This effect can be seen verifying that all the upper curves of 
Fig. 3 exhibit higher oscillation frequency in comparison with the
lower ones (corresponding to lower v$_{x}$ values).

Another parameter is the time of collapse of oscillations $\left(
t_{c}\right) $, that is, the time in which the oscillations associated with
different values of $n$ become uncorrelated. For an initial coherent state
in the cavity with sufficiently large $\bar{n}$ (for photon number standard
deviation $\Delta n=\sqrt{\bar{n}}\ll \bar{n}$), we can estimate $t_{c}$ as
(see Ref. \cite{scully}): $(\Omega _{\bar{n}+\sqrt{\bar{n}}}-\Omega _{\bar{n}%
-\sqrt{\bar{n}}})t_{c}\sim 1$. After using Eq. (\ref{O_lv}), we write 
\begin{equation}
t_{c}\sim t_{c}^{0}\left( 1+\frac{\alpha ^{2}}{2\bar{n}}\right) ,
\label{eqtc}
\end{equation}%
where $t_{c}^{0}=1/(2g)$ is the collapse time in the usual two-level quantum
system (without LV). This expression shows an enlargement of the collapse
time with the background magnitude with fixed $\bar{n}$. Indeed, the upper
curves of Fig. 1 show longer collapse times when compared with the lower
ones. Also, when comparing Figs. 1 and 3 for the same background
values, we verify that the collapse time decreases with an increasing value
of $\ \bar{n}$. This is consistent with Eq. (\ref{eqtc}). This effect is
similar to the\ collapse time behavior observed in the usual \ two-level
quantum system (without LV) at a non-resonant regime $\left( \Delta \neq
0\right) ,$ which also diminishes with $\bar{n}$ (see ref. \cite{scully}).

Finally, we regard the time of revival of oscillations, $\left( t_{r}\right) 
$, given by the condition $(\Omega _{\bar{n}}-\Omega _{\bar{n}-1})t_{r}=2\pi
m,\,\,m=1,2,...$, as the time in which two oscillators with
neighboring photon numbers $n=\bar{n}-1$ and $n=\bar{n}$ acquire a $2m\pi $
phase difference \cite{yoo}. This gives 
\begin{equation}
t_{r}\sim {t_{r}^{0}}\biggl(1+\frac{\alpha ^{2}}{2\bar{n}}\biggr)
\label{eqtr}
\end{equation}%
with $t_{r}^{0}=2\pi m\sqrt{\bar{n}}/g,$ $m=1,2,3,...$. This result turns
clear that for a fixed background the revivals take place at regular
intervals as in the usual two-level quantum theory. Also, such intervals are
augmented for increasing values of the background, becoming the revival
packages more distant (in time) from each other. This is compatible with the
behavior exhibited by the upper curves in Figs. 1 and 3.

\section{Conclusion}

In this work, we have considered the main consequences of the LV vector
coupling term $\left( v_{\mu }\overline{\psi }\gamma ^{\mu }\psi \right) $
on a quantized two-level atom coupled with a quantized electromagnetic field
in a cavity. We have written the nonrelativistic LV corrections in the
interaction picture and considered such contributions into the Schr\"{o}%
dinger equation in order to obtain the modified system of coupled
differential equations for the probability amplitude coefficients that
describe the atom-field wave function.\ Experimental values of the relevant
parameters of the model revealed that the term $\mathbf{A}\cdot \mathbf{v}$
is of lower magnitude when compared with the term $\mathbf{p}\cdot \mathbf{v}
$, which corroborates recent results from the semiclassical theory. The
Lorentz-violating Hamiltonian stemming from the quantized vector potential
has a ``non-conserving" energy character. This kind of non-conserving terms
are usually discarded in the rotating wave approximation, but here are the
ones that implied physical effects on the system.

After decoupling the system of differential equations, the probability
amplitude coefficients fulfill typical harmonic oscillator equations, whose
solutions lead to modified expressions for the population inversion function
and for the Rabi frequency. These results allow to note that the photon
statistics and the average number of photons in the cavity are not changed.
On the other hand, the Rabi frequency increases with the background value,
which is associated with a decreasing in the Rabi period $\left(
t_{R}\right) $. At the same time, the modified population inversion function
revealed that the revivals tend to occur later as the background magnitude
increases. As a consequence of the alteration implied on the Rabi frequency,
the collapse $\left( t_{c}\right) $ and revival $\left( t_{r}\right) $ times
increase with an increasing background, so that the revival packages become 
larger and more distant from each other. In order to keep these
modifications in an undetectable scale for the parameter ranges considered,
an upper bound $\left( \text{v}_{x}\leq 10^{-10}eV\right) $ for the
background was set up. Such bound is in agreement with a recent result
obtained by us from a semiclassical approach \cite{Semi}.

Note also that the deduced expressions for the modified times $t_{R},t_{c}$
and $t_{r}$ are better verified for $\bar{n}>>\sqrt{\bar{n}}$. In our
example, we have chosen $\bar{n}=25$ and $\bar{n}=5$, which does not fulfill
this condition. However, these not so large $\bar{n}$ values have already
obeyed qualitatively the tendency stated by Eqs. (\ref{eqtR}), (\ref{eqtc})
and (\ref{eqtr}). Despite a choice of \ larger values of $\bar{n}$ 
has implied a longer sequence of collapses and revivals, our
analysis showed that smaller values of $\bar{n}$ are more efficient in the
task of establishing a more stringent upper bound on the background
magnitude. 
%%%%%%%%%%%%%%%%%%%%%%%%%%%%%%%%%%%%%%%%%%%%%%%%%%%%%%%%%%%%%%%%%%%%%%%%%%%%%%%%%%%%%%%%%%%%%%%%%%%%%%%%%%%%%%%%%%%%

Some interesting issues may be regarded in forthcoming investigations, as
the examination of the spontaneous emission in the presence of Lorentz
violation, with the evaluation of LV corrections on the decaying rate.

\section*{Acknowledgments}

The authors thank CNPq, FAPEMA (brazilian agencies), and CNPq-MCT-CT-Energ
for financial support. They are also grateful to K. Furuya for Ref. \cite%
{rwa} and B. Baseia for relevant discussions.

%%%%%%%%%%%%%%%%%%%%%%%%%%%%%%%%%%%%%%%%%%

\widetext


\begin{thebibliography}{99}
\bibitem{foot1} The interaction picture may be defined by the following
equations: $|\Phi _{I}(t)\rangle =e^{iH_{0S}t/\hbar }|\Phi _{S}(t)\rangle ,$ 
$\widehat{A}_{I}(t)=e^{iH_{0S}t/\hbar }\widehat{A}_{S}e^{-iH_{0S}t/\hbar },$
where $H_{0S}$ is the Hamiltonian $H_{0}$ in the Schr\"{o}dinger
representation, $\widehat{A}_{I},\widehat{A}_{S}$ are the same operator $%
\widehat{A}$ written in the interactions and Schr\"{o}dinger pictures. These
two definitions yield the equations: $i\hbar \frac{d}{dt}|\Phi
_{I}(t)\rangle =H_{1}|\Phi _{I}(t)\rangle ,$ $\frac{d}{dt}\widehat{A}_{I}(t)=%
\frac{1}{i\hslash }[\widehat{A}_{I},H_{0I}].$
\end{thebibliography}
\end{document}